\documentclass{article}
\usepackage{amssymb}
\usepackage{amsmath}
\usepackage{geometry}

\begin{document}

\title{Quantization of soluble classical constrained systems}
\author{{\small Z. Belhadi}$^{a,b}${\small , F. Menas}$^{a,c}${\small , A. B%
\'{e}rard}$^{d}${\small \ and H. Mohrbach}$^{d}${\small .} \and \textit{a}%
{\small \ - Laboratoire de physique et chimie quantique, Facult\'{e} des
sciences. } \and {\small Universit\'{e} Mouloud Mammeri, BP 17, 15000 Tizi
Ouzou, Alg\'{e}rie.} \and \textit{b}{\small \ - Laboratoire de physique th%
\'{e}orique, Facult\'{e} des sciences exactes.} \and {\small Universit\'{e}
de Bejaia. 06000 Bejaia, Alg\'{e}rie.} \and \textit{c}{\small \ - Ecole
Nationale Pr\'{e}paratoire aux Etudes d'ing\'{e}niorat. Laboratoire de
physique.} \and {\small RN 5 Rouiba, Alger. Alg\'{e}rie.} \and \textit{d}%
{\small \ - Equipe BioPhysStat, Laboratoire LCP-A2MC, ICPMB, IF\ CNRS\ N}$%
{{}^\circ}%
2843$ \and {\small Universit\'{e} de Lorraine. 1 Bd Arago, 57078 Metz Cedex,
France.}}
\maketitle

\begin{abstract}
The derivation of the brackets among coordinates and momenta for classical
constrained systems is a necessary step toward their quantization. Here we
present a new approach for the determination of the classical brackets which
does neither require Dirac's formalism nor the symplectic method of Faddeev
and Jackiw. This approach is based on the computation of the brackets
between the constants of integration of the exact solutions of the equations
of motion. From them all brackets of the dynamical variables of the system
can be deduced in a straightforward way.
\end{abstract}

\section{Introduction}

The Hamiltonian method for the quantization of a dynamical system requires
to postulate canonical Poisson brackets among coordinates and momenta. Then
the correspondence principle is applied to determine the different
commutators of the quantum operators associated to the classical dynamical
variables. However, for singular Lagrangians or constrained systems, this
procedure does not work and the brackets have to be determined and not
postulated. Dirac has developed a general formalism for treating such
systems \cite{Dirac,Bergmann}. Dirac formalism has been widely used but the
determination of the Hamiltonian and the Dirac brackets replacing the
Poisson brackets can be very cumbersome. More recently Faddeev and Jackiw
have proposed \cite{FJ,LL} an alternative approach (FJ approach) based on
the symplectic formalism and Darboux theorem (which is equivalent to an
approach proposed by Souriau \cite{Souriau} as remarked in \cite{Horvathy}
and \cite{Martina} ) to find the brackets and the Hamiltonian which can
often avoid many of the steps of Dirac method.

Here we propose another approach for the determination of the Hamiltonian
and the brackets for unconstrained and constrained systems which avoids the
Dirac categorization of constraints and the Darboux theorem of the FJ
approach. Our method is based on the computation of the brackets between the
constants of integration of the solutions of the Euler-Lagrange equations of
motion. From them all brackets among coordinates and momenta can be directly
deduced. Therefore when a system is classically soluble, i.e the analytical
solution of its equations of motion are known, it is possible to quantize
the system without using Dirac or FJ formalisms.

In the two next sections, the general formalism for the computation of the
brackets among the constants of integration and the link with Dirac and FJ
methods is presented. Then in the following sections we provide several
examples which are treated with the three methods. At the end of the paper
we mention possible applications to field theory.

\section{\textbf{Remark on the integration constants}}

Consider a system of $N$ degrees of freedom characterized by its generalized
coordinates $q=(q_{1},q_{2},...,q_{N})$ and conjugate momenta $%
p=(p_{1},p_{2},...,p_{N})$. Now let $f=f(q,p)$ and $g=g(q,p)$ be two
arbitrary functions. We define an antisymmetric bilinear bracket satisfying
the Leibniz rule and the Jacobi identities, such that:%
\begin{equation}
\left\{ f,g\right\} =\sum_{i,j=1}^{N}\left( \left\{ q_{i},q_{j}\right\} 
\frac{\partial f}{\partial q_{i}}\frac{\partial g}{\partial q_{j}}+\left\{
p_{i},p_{j}\right\} \frac{\partial f}{\partial p_{i}}\frac{\partial g}{%
\partial p_{j}}+\left\{ q_{i},p_{j}\right\} \left( \frac{\partial f}{%
\partial q_{i}}\frac{\partial g}{\partial p_{j}}-\frac{\partial f}{\partial
p_{j}}\frac{\partial g}{\partial q_{i}}\right) \right)  \label{hjkllm}
\end{equation}%
At this level this bracket is not yet specified. It can be either the
Poisson, or Dirac or also a FJ bracket$.$ Now, suppose that $q$ and $p$ are
functions of time and of new variables $R=(R_{1},R_{2},...,R_{M})$ with $%
M\leq 2N$, i.e \footnote{$M=2N$ corresponds to an unconstrained system. For
the constrained systems, each constraint eliminates one variable.} $q=q(R,t)$
and $p=p(R,t)$. Then Eq. (\ref{hjkllm}) becomes%
\begin{eqnarray*}
\{f,g\} &=&\sum_{i,j=1}^{N}\sum_{k,l=1}^{M}\left( \{q_{i},q_{j}\}\frac{%
\partial f}{\partial R_{k}}\frac{\partial R_{k}}{\partial q_{i}}\frac{%
\partial g}{\partial R_{l}}\frac{\partial R_{l}}{\partial q_{j}}%
+\{p_{i},p_{j}\}\frac{\partial f}{\partial R_{k}}\frac{\partial R_{k}}{%
\partial p_{i}}\frac{\partial g}{\partial R_{l}}\frac{\partial R_{l}}{%
\partial p_{j}}\right) \\
&&+\sum_{i,j=1}^{N}\sum_{k,l=1}^{M}\{q_{i},p_{j}\}\left( \frac{\partial f}{%
\partial R_{k}}\frac{\partial R_{k}}{\partial q_{i}}\frac{\partial g}{%
\partial R_{l}}\frac{\partial R_{l}}{\partial p_{j}}-\frac{\partial f}{%
\partial R_{k}}\frac{\partial R_{k}}{\partial q_{j}}\frac{\partial g}{%
\partial R_{l}}\frac{\partial R_{l}}{\partial p_{i}}\right)
\end{eqnarray*}%
which simplifies to 
\begin{eqnarray*}
\{f,g\} &=&\sum_{k,l=1}^{M}\left[ \sum_{i,j=1}^{N}\{q_{i},q_{j}\}\frac{%
\partial R_{k}}{\partial q_{i}}\frac{\partial R_{l}}{\partial q_{j}}%
+\sum_{i,j=1}^{N}\{p_{i},p_{j}\}\frac{\partial R_{k}}{\partial p_{i}}\frac{%
\partial R_{l}}{\partial p_{j}}\right. \\
&&+\left. \sum_{i,j=1}^{N}\{q_{i},p_{j}\}\left( \frac{\partial R_{k}}{%
\partial q_{i}}\frac{\partial R_{l}}{\partial p_{j}}-\frac{\partial R_{k}}{%
\partial q_{j}}\frac{\partial R_{l}}{\partial p_{i}}\right) \right] \frac{%
\partial f}{\partial R_{k}}\frac{\partial g}{\partial R_{l}}
\end{eqnarray*}%
and leads to the expression of $\left\{ f,g\right\} $ in terms of $\left\{
R_{k},R_{l}\right\} $:

\begin{equation}
\left\{ f,g\right\} =\sum_{k,l=1}^{M}\left\{ R_{k},R_{l}\right\} \frac{%
\partial f}{\partial R_{k}}\frac{\partial g}{\partial R_{l}}
\label{EqFondamental}
\end{equation}%
With the introduction of the new variables, the sum goes from $1$ to $M$ ($%
M\leq 2N$)$.$ Thus, the presence of constraints reduces the number of
variables and we will only work with the free variables. Now, if $f=q_{i}$
or $p_{i}$ and $g=H$ the Hamiltonian of the system, we obtain the following
equations (using Hamilton equations for $i=1$ to $N$)

\begin{eqnarray}
\frac{dq_{i}}{dt} &=&\left\{ q_{i},H\right\} \text{ \ \ \ }\Rightarrow \text{
\ \ \ \ \ }\frac{\partial q_{i}}{\partial t}+\sum_{j=1}^{M}\frac{\partial
q_{i}}{\partial R_{j}}\frac{dR_{j}}{dt}=\sum_{j,k=1}^{M}\left\{
R_{j},R_{k}\right\} \frac{\partial q_{i}}{\partial R_{j}}\frac{\partial H}{%
\partial R_{k}}  \notag \\
\frac{dp_{i}}{dt} &=&\left\{ p_{i},H\right\} \text{ \ \ \ }\Rightarrow \text{
\ \ \ \ \ }\frac{\partial p_{i}}{\partial t}+\sum_{j=1}^{M}\frac{\partial
p_{i}}{\partial R_{j}}\frac{dR_{j}}{dt}=\sum_{j,k=1}^{M}\left\{
R_{j},R_{k}\right\} \frac{\partial p_{i}}{\partial R_{j}}\frac{\partial H}{%
\partial R_{k}}  \label{zahir}
\end{eqnarray}%
Now suppose that\textbf{\ }$R_{k},k=1...M$\textbf{\ }are constants of motion
(first integrals)\textbf{, }denoted\textbf{\ }$R_{k}=C_{k},k=1...M,$\textbf{%
\ }then the above equations will be reduced to the form%
\begin{eqnarray}
\frac{\partial q_{i}}{\partial t} &=&\sum_{j,k=1}^{M}\left\{
C_{j},C_{k}\right\} \frac{\partial q_{i}}{\partial C_{j}}\frac{\partial H}{%
\partial C_{k}}  \label{zahir2} \\
\frac{\partial p_{i}}{\partial t} &=&\sum_{j,k=1}^{M}\left\{
C_{j},C_{k}\right\} \frac{\partial p_{i}}{\partial C_{j}}\frac{\partial H}{%
\partial C_{k}}\text{ \ \ \ \ \ }i=1...N
\end{eqnarray}%
Therefore the dynamics of the system can be expressed in terms of the
brackets among the integration constants, without ever specifying the nature
of the bracket (Poisson, Dirac or FJ brackets ). Therefore the relations Eqs.%
$\left( \ref{zahir2}\right) $ are universal and common to all types of
brackets. With the introduction of the notation\textbf{\ }$\left. \xi
_{i}\right\vert _{i=1,...2N}=(q_{i},p_{i})$ we can now write Eqs.$\left( \ref%
{zahir2}\right) $ in a compact form :

\begin{equation}
\frac{\partial \xi _{i}}{\partial t}=\sum_{j,k=1}^{M}\{C_{j},C_{k}\}\frac{%
\partial \xi _{i}}{\partial C_{j}}\frac{\partial H}{\partial C_{k}}\ \ \ \ \
\ i=1...2N  \label{FJ equivalent}
\end{equation}%
These relations can obtained in the same manner as we did by starting from
the equations of motion 
\begin{equation*}
\overset{.}{\xi }_{i}=\left\{ \xi _{i},H\right\} =\sum_{j=1}^{2N}\left\{ \xi
_{i},\xi _{j}\right\} \frac{\partial H}{\partial \xi _{j}}\ \ \ \ \ \
i=1...2N
\end{equation*}%
used in the FJ approach.

\section{Method of Integration constants: Constrained (and unconstrained)
systems}

In this section we use the previous properties Eqs. $\left( \ref{zahir2}%
\right) $ to determine the brackets of any exactly soluble system. At the
end the canonical quantization allows us to build the quantum version.
Although our approach (called in the rest of the paper the CI method) is
applicable to systems without constraints, it is for constrained systems
that it shows its main interest. Indeed, we propose a new method which
greatly facilitates the computation of the Dirac brackets, while avoiding
both the complicated Dirac algorithm and the Darboux theorem in the case of
the FJ approach.

Consider a classical system described by a singular autonomous Lagrangian $%
L(q,\dot{q})$ where $q=(q_{1},...,q_{N})$ are the generalized coordinates
and $\dot{q}=(\dot{q}_{1},...,\dot{q}_{N})$ the generalized velocities.
Suppose we know the (general) analytical solutions $q(t)=\tilde{q}(t,C)$ of
the Euler-Lagrange equations and the momenta$\ p(t)=\tilde{p}(t,C)$ $\left(
p_{i}=\frac{\partial L}{\partial \dot{x}_{i}}\right) $, where $%
C=(C_{1},C_{2},...,C_{M})$ is the set of constants of integration. For
constrained systems we have obviously $M<2N.$

Before going further, it is necessary to distinguish between two cases: the
first is when there are no arbitrary functions in the solutions (no gauge
symmetry), the second is otherwise. In the later case with a gauge symmetry,
we must first choose the arbitrary functions once by adding new conditions
(fixing the gauge) before moving to the canonical formalism and defining any
brackets.

From the analytical solutions of the equations of motion we can write the
Hamiltonian\footnote{%
We can also obtain the Hamiltonian by putting the solutions into the
Legendre transformation $H=\sum_{i}\frac{d\tilde{q}_{i}(t,C)}{dt}\tilde{p}%
_{i}(t,C)-L\left( \tilde{q}_{i}(t,C),\frac{d\tilde{q}_{i}(t,C)}{dt}\right) .$
In this way we do not have to inverse the momenta with respect to the
velocities.} as $H(q(t),p(t))=H(\tilde{q}(t,C),\tilde{p}(t,C))$ and from
Eqs. $\left( \ref{zahir}\right) $ we deduce the fundamental equations

\begin{eqnarray}
\frac{\partial }{\partial t}\tilde{q}_{i}(t,C)
&=&\sum_{j,k=1}^{M}\{C_{j},C_{k}\}\frac{\partial \tilde{q}_{i}}{\partial
C_{j}}\frac{\partial H}{\partial C_{k}}\ \ \ \ i=1...N  \notag \\
\frac{\partial }{\partial t}\tilde{p}_{i}(t,C)
&=&\sum_{j,k=1}^{M}\{C_{j},C_{k}\}\frac{\partial \tilde{p}_{i}}{\partial
C_{j}}\frac{\partial H}{\partial C_{k}}\ \ \ \ i=1...N  \label{eq of motion}
\end{eqnarray}%
These $2N$ equations contain $M(M-1)/2$ unknown brackets $\{C_{j},C_{k}\},$
with $j,k=1...M$. Our method consist in determining the brackets $%
\{C_{j},C_{k}\}$ from the Eqs.$\left( \ref{eq of motion}\right) $ via a
simple identification. But, with this procedure, we easily see that only the
brackets containing at least one integration constant in the expression of
the Hamiltonian are available. To solve this problem, we must add
supplementary terms to the Lagrangian. In other words, if for example the
bracket $\{C_{i},C_{j}\}$ is not accessible and if one of these constants
appears in the expression of the generalized coordinate $q_{i}$, one has to
add a term of the form $\eta q_{i}$\ to the Lagrangian, redo all the
calculations and put $\eta =0$ at the end (as an illustration see example
4.2)$.$ Using the brackets $\{C_{i},C_{j}\}$ we can compute the brackets $%
\left\{ q_{i},q_{j}\right\} $, $\left\{ p_{i},p_{j}\right\} $ and $\left\{
q_{i},p_{j}\right\} $ more easily than with any other existing approaches$.$
If the result of the calculation depends on the integration constants, it is
possible to make them disappear by inverting the solutions $\tilde{q}(t,C)$
and $\tilde{p}(t,C)$. At this time, to be sure of the validity of our
calculations, we can just verify that $\tilde{q}(t,C)$ and $\tilde{p}(t,C)$
are solutions of the equations of motion obtained from the Hamilton
equations using these fundamental brackets. We see that in our method we do
not even talk about constraints unlike other approaches.

These brackets are essential for the quantization of the system by
introducing the quantum operators $\widehat{q}$ and $\widehat{p}$ and the
principle of correspondence we have : 
\begin{equation}
\left[ f(\hat{q},\hat{p}),g(\hat{q},\hat{p})\right] =i\hbar \widehat{%
\{f(q,p),g(q,p)\}}
\end{equation}

Note, that it might look surprising to see brackets between constants and
the derivative of functions with respect to those constants. Actually, these
constants are first integrals and must be considered as in the
Hamilton-Jacobi formalism which also treats the constants as variables.

\section{Applications}

In this section we present several appealing examples. As a first
application of the method we consider the isotonic oscillator which is an
unconstrained system. This case shows that the method is general and thus
valid for constrained and unconstrained systems. The second application
deals with a singular autonomous Lagrangian which exemplifies the previous
discussion about the procedure when $\{C_{i},C_{j}\}$ is not directly
accessible (when the Hamiltonian does not contain $C_{i}$ or $C_{j}$). The
other examples concern different constrained systems which are studied with
the Dirac, FJ and our method for comparison.

\subsection{Regular Lagrangian}

For the first example we consider an unconstrained system namely the
Lagrangian of the isotonic (anharmonic) oscillator: 
\begin{equation*}
L=\frac{1}{2}\dot{x}^{2}-\frac{1}{2}\omega ^{2}x^{2}-\frac{k}{x^{2}}
\end{equation*}%
The Euler-Lagrange equations lead to the following non-linear equation 
\begin{equation*}
\ddot{x}+\omega ^{2}x-\frac{2k}{x^{3}}=0
\end{equation*}%
whose solution is (see ref: \cite{f}\textbf{) }$x(t)=\frac{1}{A\omega }\sqrt{%
(A^{4}\omega ^{2}-2k)\sin ^{2}(wt+\phi )+2k}$\textbf{\ }where $A$\ and $\phi 
$\ are the constants of integration. The conjugate momentum is given by $%
p_{x}(t)=\dot{x}(t)=\frac{(A^{4}\omega ^{2}-2k)\sin (wt+\phi )\cos (wt+\phi )%
}{A\sqrt{(A^{4}\omega ^{2}-2k)\sin ^{2}(wt+\phi )+2k}}.$ Expressing the
Hamiltonian $H=\frac{1}{2}p_{x}^{2}+\frac{1}{2}\omega ^{2}x^{2}+\frac{k}{%
x^{2}}$ in terms of these constants we obtain

\begin{equation*}
H=\frac{A^{4}\omega ^{2}+2k}{2A^{2}}
\end{equation*}%
From the Hamilton equation $\dot{x}=\{x,H\}$ and the property Eq. \textbf{(}%
\ref{eq of motion}) we obtain $\frac{\partial x}{\partial t}=\{\phi ,A\}%
\frac{\partial x}{\partial \phi }\frac{\partial H}{\partial A}$ from which
we obtain the bracket between the constants of integration\textbf{\ }%
\begin{equation*}
\{\phi ,A\}=\frac{A^{3}\omega }{A^{4}\omega ^{2}-2k}
\end{equation*}%
From Eq. \textbf{(\ref{EqFondamental}) we }obtain\ the brackets among the
dynamical variables: 
\begin{equation*}
\{x,p_{x}\}=\{\phi ,A\}\left( -\frac{\partial x}{\partial A}\frac{\partial
p_{x}}{\partial \phi }+\frac{\partial x}{\partial \phi }\frac{\partial p_{x}%
}{\partial A}\right)
\end{equation*}%
A short computation shows that $-\frac{\partial x}{\partial A}\frac{\partial
p_{x}}{\partial \phi }+\frac{\partial x}{\partial \phi }\frac{\partial p_{x}%
}{\partial A}=\frac{A^{4}\omega ^{2}-2k}{A^{3}\omega }$ so that we retrieve
the canonical Poisson bracket 
\begin{equation*}
\{x,p_{x}\}=1
\end{equation*}%
as expected for an unconstrained system.

\subsection{Singular autonomous Lagrangians: first example}

Consider a constrained system described by the Lagrangian : 
\begin{equation*}
L=\frac{\dot{x}^{2}}{2}+x\dot{y}-y\dot{z}
\end{equation*}%
This system gives an illustration of the discussed procedure when the
Hamiltonian does not contain all constants of integration. The
Euler-Lagrange equations are 
\begin{eqnarray*}
\ddot{x}-\dot{y} &=&0 \\
\dot{x}+\dot{z} &=&0 \\
\dot{y} &=&0
\end{eqnarray*}%
whose analytical solutions are : 
\begin{equation*}
x(t)=at+b\ \ \ \ y(t)=c\ \ \ \ z(t)=-at+d
\end{equation*}%
where $a,b,c$ and $d$ are the constants of integration. From the conjugate
momenta $p_{x}=\dot{x},\ p_{y}=x$\textbf{\ }and$\ p_{z}=-y,$ and the
Legendre transformation we obtain the Hamiltonian

\begin{equation*}
H=\frac{p_{x}^{2}}{2}=\frac{a^{2}}{2}
\end{equation*}%
From this Hamiltonian only the brackets containing $a$ are accessible. To
solve this issue we introduce a new Lagrangian%
\begin{equation*}
L=\frac{\dot{x}^{2}}{2}+x\dot{y}-y\dot{z}-\lambda x-\xi y
\end{equation*}%
where $\lambda $ and $\xi $ are real parameters. The new equations of motion
are thus $\ddot{x}-\dot{y}+\lambda =0,\ \dot{x}+\dot{z}+\xi =0$ and $\dot{y}%
=0$\textbf{.} The general solution of this system reads

\begin{equation}
x(t)=-\frac{\lambda }{2}t^{2}+at+b\ \ \ \ \ \ \ \ \ \ y(t)=c\ \ \ \ \ \ \ \
\ z(t)=\frac{\lambda }{2}t^{2}-at-\xi t+d  \label{dffrr}
\end{equation}%
The Hamiltonian now reads\textbf{\ }$H=\frac{p_{x}^{2}}{2}+\lambda x+\xi y$%
\textbf{\ }or in terms of the constants of integration 
\begin{equation*}
H=\frac{1}{2}a^{2}+\lambda b+\xi c
\end{equation*}%
From the Hamilton equations we can directly obtain the brackets between the
different constants of integration. Indeed we have 
\begin{equation*}
\dot{x}=\{x,H\}\Rightarrow -\lambda t+a=t\lambda \{a,b\}+\xi
t\{a,c\}+a\{b,a\}+\xi \{b,c\}
\end{equation*}%
and 
\begin{equation*}
\overset{\cdot }{z}=\{z,H\}\Rightarrow \lambda t-a-\xi =-t\lambda
\{a,b\}-\xi t\{a,c\}+a\{d,a\}+\lambda t\{d,b\}+\xi \{d,c\}
\end{equation*}%
which by identification gives:

\begin{eqnarray*}
\{a,b\} &=&\{a,d\}=-1\text{ } \\
\{c,d\} &=&1 \\
\{a,c\} &=&\{b,c\}=\{b,d\}=0\text{ }
\end{eqnarray*}%
A direct calculation using these results and the solution Eq. (\ref{dffrr})
leads the following brackets among the dynamical variables\textbf{\ }%
\begin{eqnarray*}
\{x,p_{x}\} &=&\{y,z\}=\{z,p_{z}\}=1\text{ } \\
\{p_{x},p_{y}\}\text{ } &=&\{p_{x},z\}=-1
\end{eqnarray*}%
These brackets do not depend on the parameters $\lambda $ and $\xi $ and are
thus unchanged if we put $\lambda =\xi =0.$ In other words they are also the
brackets of the initial Lagrangian\textbf{\ }$L=\frac{\dot{x}^{2}}{2}+x\dot{y%
}-y\dot{z}-\lambda x-\xi y|_{\lambda =\xi =0}=\frac{\dot{x}^{2}}{2}+x\dot{y}%
-y\dot{z}.$

\subsection{Singular autonomous Lagrangians: second example}

Consider a system described by the Lagrangian : 
\begin{equation*}
L=\frac{\dot{x}^{2}}{2}+\frac{x^{2}}{2}\dot{y}-\frac{x^{2}}{2}y
\end{equation*}

\subsubsection{CI\ method}

The Euler-Lagrange equations are 
\begin{eqnarray*}
\ddot{x} &=&x\dot{y}-xy\ \ \ \ \ \ \ \ \ \ x\dot{x}=\frac{x^{2}}{2} \\
p_{x} &=&\dot{x}\ \ \ \ \ \ \ \ \ \ \ \ \ \ \ \ \ \ \ p_{y}=\frac{x^{2}}{2}
\end{eqnarray*}%
whose analytical solutions are :

\begin{eqnarray*}
x(t) &=&ae^{-\frac{1}{2}t}\ \ \ \ \ \ \ \ \ \ \ \ \ \ \ \ \ \ \ \ \ \ \ \ \
\ \ \ \ y(t)=be^{t}-\frac{1}{4} \\
p_{x}(t) &=&-\frac{a}{2}e^{-\frac{1}{2}t}=-\frac{x}{2}\ \ \ \ \ \ \ \ \ \ \
\ \ \ \ \ p_{y}(t)=\frac{a^{2}}{2}e^{-t}
\end{eqnarray*}%
where $a$ and $b$ are the constants of integration. Expressing the
Hamiltonian in terms of these constants we have 
\begin{equation*}
H=\left( p_{x}^{2}+x^{2}y\right) /2=a^{2}b/2
\end{equation*}%
From Eqs.$\left( \text{\ref{eq of motion}}\right) $ we have for the variable 
$x$,

\begin{equation*}
\dot{x}=\left\{ x,H\right\} \text{ \ }\Rightarrow \text{ \ }\frac{a}{2}e^{-%
\frac{1}{2}t}=-\{ae^{-\frac{1}{2}t},\frac{a^{2}b}{2}\}\text{ \ }\Rightarrow 
\text{ \ }\frac{a}{2}e^{-\frac{1}{2}t}=-e^{-\frac{1}{2}t}\frac{a^{2}}{2}%
\{a,b\}
\end{equation*}%
and then by identification%
\begin{equation*}
\{a,b\}=\frac{-1}{a}
\end{equation*}%
From this brackets we obtain directly all other brackets among the dynamical
variables. For instance:

\begin{equation*}
\{x,y\}=\{ae^{-\frac{1}{2}t},be^{t}-\frac{1}{4}\}=e^{\frac{1}{2}t}\{a,b\}=%
\frac{-1}{a}e^{\frac{1}{2}t}=-\frac{1}{x}.
\end{equation*}%
The other brackets are calculated in the same way, leading to 
\begin{eqnarray*}
\{x,p_{x}\} &=&\{x,p_{y}\}=\{p_{x},p_{y}\}=0 \\
\{y,p_{x}\} &=&-\frac{1}{2x} \\
\{y,p_{y}\} &=&1
\end{eqnarray*}%
To check the validity of the results we determine the Hamilton equations by
using the derived brackets, which are: 
\begin{eqnarray*}
\dot{x} &=&\{x,H\}\Rightarrow \dot{x}=-x/2\text{ \ ; \ }\dot{y}%
=\{y,H\}\Rightarrow \dot{y}=-p_{x}/(2x)+y \\
\dot{p}_{x} &=&\{p_{x},H\}\Rightarrow \dot{p}_{x}=x/4\text{ ; }\dot{p}%
_{y}=\{p_{y},H\}\Rightarrow \dot{p}_{y}=-x^{2}/2
\end{eqnarray*}%
These equations are equivalent to the Euler-Lagrange equations, which
validates our approach for this first example.

\subsubsection{Dirac\ approach}

For comparison we consider the same Lagrangian with the Dirac treatment.
First we deduce the momenta $p_{x}=\dot{x}$ and $p_{y}=\frac{x^{2}}{2}$ and
the canonical Hamiltonian 
\begin{equation*}
H_{c}=\frac{p_{x}^{2}}{2}+\frac{x^{2}}{2}y.
\end{equation*}%
As we have a primary constraint $\phi =p_{y}-\frac{x^{2}}{2}=0$, we
introduce the total Hamiltonian $H_{T}=H_{c}+\lambda \phi $ where $\lambda $

is a Lagrange multiplier. The consistency condition $\dot{\phi}\approx
\left\{ \phi ,H_{T}\right\} _{PB}\approx \left\{ \phi ,H_{c}\right\}
_{PB}\approx 0$ (where $\left\{ ,\right\} _{PB}$ denotes the Poisson
bracket) leads to a secondary constraint $\zeta =\frac{x}{2}+p_{x}\approx 0$%
. The associated consistency condition $\dot{\zeta}\approx \left\{ \zeta
,H_{c}\right\} _{PB}+\lambda \left\{ \zeta ,\phi \right\} _{PB}\approx 0$
leads us to the following expression $\lambda \approx y-\frac{p_{x}}{2x}.$

As $\{\phi ,\zeta \}_{PB}\approx -x\neq 0,$ the constraints are second class
and the matrix of constraints is given by 
\begin{equation*}
M=\left( \hspace{0cm}%
\begin{tabular}{ll}
$\left\{ \phi ,\phi \right\} _{PB}$ & $\hspace{0cm}\left\{ \phi ,\zeta
\right\} _{PB}$ \\ 
$\left\{ \zeta ,\phi \right\} _{PB}$ & $\hspace{0cm}\left\{ \zeta ,\zeta
\right\} _{PB}$%
\end{tabular}%
\right) =\left( \hspace{0cm}%
\begin{tabular}{ll}
$0$ & $-x$ \\ 
$x$ & $\hspace{0cm}0$%
\end{tabular}%
\right) .
\end{equation*}%
From the different elements of the inverse matrix

\begin{equation*}
M^{-1}=\left( \hspace{0cm}%
\begin{tabular}{ll}
$\left\{ \phi ,\phi \right\} _{PB}^{-1}$ & $\hspace{0cm}\left\{ \phi ,\zeta
\right\} _{PB}^{-1}$ \\ 
$\left\{ \zeta ,\phi \right\} _{PB}^{-1}$ & $\hspace{0cm}\left\{ \zeta
,\zeta \right\} _{PB}^{-1}$%
\end{tabular}%
\right) =\left( \hspace{0cm}%
\begin{tabular}{ll}
$0$ & $\hspace{0cm}\frac{1}{x}$ \\ 
$-\frac{1}{x}$ & $\hspace{0cm}0$%
\end{tabular}%
\right)
\end{equation*}%
and those of the matrix of constraints, we can obtain the Dirac bracket $%
\left\{ f,g\right\} _{D}$ among arbitrary functions which is defined as

\begin{equation*}
\left\{ f,g\right\} _{D}=\left\{ f,g\right\} _{PB}-\left\{ f,\phi
_{1}\right\} _{PB}\left\{ \phi _{1},\phi _{2}\right\} _{PB}^{-1}\left\{ \phi
_{2},g\right\} _{PB}-\left\{ f,\phi _{2}\right\} _{PB}\left\{ \phi _{2},\phi
_{1}\right\} _{PB}^{-1}\left\{ \phi _{1},g\right\} _{PB}
\end{equation*}%
Applying this formula to the coordinates and momenta we get the following
brackets:

\begin{eqnarray*}
\left\{ x,y\right\} _{D} &=&\left\{ x,y\right\} _{PB}-\left\{ x,\phi
\right\} _{PB}\frac{1}{x}\left\{ \zeta ,y\right\} _{PB}+\left\{ x,\zeta
\right\} _{PB}\frac{1}{x}\left\{ \phi ,y\right\} _{PB}=-\frac{1}{x} \\
\left\{ x,p_{x}\right\} _{D} &=&\left\{ x,p_{x}\right\} _{PB}-\left\{ x,\phi
\right\} _{PB}\frac{1}{x}\left\{ \zeta ,p_{x}\right\} _{PB}+\left\{ x,\zeta
\right\} _{PB}\frac{1}{x}\left\{ \phi ,p_{x}\right\} _{PB}=0 \\
\left\{ x,p_{y}\right\} _{D} &=&\left\{ x,p_{y}\right\} _{PB}-\left\{ x,\phi
\right\} _{PB}\frac{1}{x}\left\{ \zeta ,p_{y}\right\} _{PB}+\left\{ x,\zeta
\right\} _{PB}\frac{1}{x}\left\{ \phi ,p_{y}\right\} _{PB}=0 \\
\left\{ y,p_{x}\right\} _{D} &=&\left\{ y,p_{x}\right\} _{PB}-\left\{ y,\phi
\right\} _{PB}\frac{1}{x}\left\{ \zeta ,p_{x}\right\} _{PB}+\left\{ y,\zeta
\right\} _{PB}\frac{1}{x}\left\{ \phi ,p_{x}\right\} _{PB}=-\frac{1}{2x} \\
\left\{ y,p_{y}\right\} _{D} &=&\left\{ y,p_{y}\right\} _{PB}-\left\{ y,\phi
\right\} _{PB}\frac{1}{x}\left\{ \zeta ,p_{y}\right\} _{PB}+\left\{ y,\zeta
\right\} _{PB}\frac{1}{x}\left\{ \phi ,p_{y}\right\} _{PB}=1 \\
\left\{ p_{x},p_{y}\right\} _{D} &=&\left\{ p_{x},p_{y}\right\}
_{PB}-\left\{ p_{x},\phi \right\} _{PB}\frac{1}{x}\left\{ \zeta
,p_{y}\right\} _{PB}+\left\{ p_{x},\zeta \right\} _{PB}\frac{1}{x}\left\{
\phi ,p_{y}\right\} _{PB}=0
\end{eqnarray*}%
which are the same as with the CI\ method.

\subsubsection{Faddeev-Jackiw approach}

It is instructive to compare our method to the FJ approach. The initial
Lagrangian is written in the following form, that we call the FJ Lagrangian%
\begin{equation*}
L_{FJ}=\dot{x}p_{x}+\dot{y}p_{y}-H=z\dot{x}+\frac{x^{2}}{2}\dot{y}-\frac{%
z^{2}}{2}-\frac{x^{2}}{2}y
\end{equation*}%
using the primary constraint $p_{y}=\frac{x^{2}}{2}$ and the notation $%
z=p_{x}.$ In this approach one has to treat $x,$ $y$ and $z$ as independent
variables. Then the Euler-Lagrange equations\footnote{%
It is useful to write the Euler-Lagrange's equations under the form $-\frac{d%
}{dt}\frac{\partial L}{\partial \dot{q}_{i}}=-\frac{\partial L}{\partial
q_{i}}$ in order to have directly the matrix $f$ antisymmetric.
\par
{}
\par
{}} are 
\begin{eqnarray*}
-\dot{z} &=&-x\dot{y}+xy \\
-x\dot{x} &=&+\frac{x^{2}}{2} \\
0 &=&-\dot{x}+z
\end{eqnarray*}%
which can be written in matrix form:%
\begin{equation*}
\underset{f}{\underbrace{\left( 
\begin{array}{ccc}
0 & x & -1 \\ 
-x & 0 & 0 \\ 
1 & 0 & 0%
\end{array}%
\right) }}\left( 
\begin{array}{c}
\dot{x} \\ 
\dot{y} \\ 
\dot{z}%
\end{array}%
\right) =\left( 
\begin{array}{c}
xy \\ 
\frac{x^{2}}{2} \\ 
z%
\end{array}%
\right)
\end{equation*}%
The matrix $f$ is singular and antisymmetric ($\det (f)=0)$. The only zero
mode is $(%
\begin{array}{ccc}
0 & 1 & x%
\end{array}%
),$ and we obtain the secondary constraint by multiplying the last equation
by this zero mode

\begin{equation*}
(%
\begin{array}{ccc}
0 & 1 & x%
\end{array}%
)\left( 
\begin{array}{c}
xy \\ 
\frac{x^{2}}{2} \\ 
z%
\end{array}%
\right) =0
\end{equation*}%
from which we obtain the constraint $\frac{x}{2}+z=0$, which is the same
secondary constraint as with Dirac approach. This constraint must be
conserved in time ($\frac{\dot{x}}{2}+\dot{z}=0$). To assure that we define
a new Lagrangian by adding the term $\lambda (\frac{\dot{x}}{2}+\dot{z})$ to
the Lagrangian $L_{FJ}$. This term can be transformed in $-\dot{\lambda}(%
\frac{x}{2}+z)$ to give the following expression

\begin{equation*}
L_{FJ}^{1}=z\dot{x}+\frac{x^{2}}{2}\dot{y}-\frac{z^{2}}{2}-\frac{x^{2}}{2}y+%
\dot{\lambda}(\frac{x}{2}+z)
\end{equation*}%
where $\lambda $ is a Lagrange multiplier which is considered as a new
variable in the following. Therefore the new Euler-Lagrange equations are

\begin{eqnarray*}
-\dot{z} &=&-x\dot{y}+xy-\left( 1/2\right) \dot{\lambda} \\
-x\dot{x} &=&\frac{x^{2}}{2} \\
0 &=&-\dot{x}+z-\dot{\lambda} \\
-(\frac{\dot{x}}{2}+\dot{z}) &=&0
\end{eqnarray*}%
which can be written in matrix form as:%
\begin{equation*}
\underset{f}{\underbrace{\left( 
\begin{array}{cccc}
0 & x & -1 & +1/2 \\ 
-x & 0 & 0 & 0 \\ 
1 & 0 & 0 & 1 \\ 
-1/2 & 0 & -1 & 0%
\end{array}%
\right) }}\left( 
\begin{array}{c}
\dot{x} \\ 
\dot{y} \\ 
\dot{z} \\ 
\dot{\lambda}%
\end{array}%
\right) =%
\begin{array}{c}
xy \\ 
\frac{x^{2}}{2} \\ 
z \\ 
0%
\end{array}%
\end{equation*}%
where $f$ is now antisymmetric and invertible as $\det (f)=x^{2}\neq 0.$ The
inverse of this matrix is : 
\begin{equation*}
f^{-1}=\left( 
\begin{array}{cccc}
\{x,x\} & \{x,y\} & \{x,z\} & \{x,\lambda \} \\ 
\{y,x\} & \{y,y\} & \{y,z\} & \{y,\lambda \} \\ 
\{z,x\} & \{z,y\} & \{z,z\} & \{z,\lambda \} \\ 
\{\lambda ,x\} & \{\lambda ,y\} & \{\lambda ,z\} & \{\lambda ,\lambda \}%
\end{array}%
\right) =\left( 
\begin{array}{cccc}
0 & -\frac{1}{x} & 0 & 0 \\ 
\frac{1}{x} & 0 & -\frac{1}{2x} & -\frac{1}{x} \\ 
0 & \frac{1}{2x} & 0 & -1 \\ 
0 & \frac{1}{x} & 1 & 0%
\end{array}%
\right)
\end{equation*}%
Then it follows that 
\begin{equation*}
\{x,y\}=-1/x\ \ \ \ ;\ \ \ \{y,z\}=-\frac{1}{2x}\ ;\text{ \ \ }\{x,z\}=0
\end{equation*}%
To have the brackets with $p_{y}$ we use the primary constraint $p_{y}=\frac{%
x^{2}}{2}.$ We \ have obtained the same result as in the case with the CI
method.\ 

\subsection{Lagrangian of a particle in a constant strong magnetic field}

We consider the simple problem of a non-relativistic particle of charge $q$
and mass $m$, in a constant magnetic field $\overrightarrow{B}_{0}$ pointing
in the $z$ direction. The Lagrangian is%
\begin{equation*}
L=\frac{1}{2}mv^{2}+q\overrightarrow{A}(x,y).\overrightarrow{v}-qV(x,y)
\end{equation*}%
where $\overrightarrow{A}\left( x,y\right) $ is the vector potential in the
Coulomb gauge $\overrightarrow{A}=\frac{1}{2}\overrightarrow{r}\wedge 
\overrightarrow{B_{0}}$ and $V(x,y)$ the scalar potential. In the limit of
the strong magnetic field the mass term can be neglected and the Lagrangian
is approximated as:

\begin{equation*}
L=\frac{qB_{0}}{2}\left( x\dot{y}-y\dot{x}\right) -qV(x,y)
\end{equation*}%
In order to simplify the mathematics we consider the case $V(x,y)=\frac{1}{2}%
k\left( x^{2}+y^{2}\right) $, thus: 
\begin{equation*}
L=\eta \left( x\dot{y}-y\dot{x}\right) -\frac{1}{2}\xi \left(
x^{2}+y^{2}\right)
\end{equation*}%
with $\eta =\frac{qB_{0}}{2}$ and $\xi =qk.$

\subsubsection{CI method}

The Euler-Lagrange equations are

\begin{equation*}
-2\eta \dot{y}+\xi x=0\text{ \ \ \ \ };\text{ \ \ \ \ }2\eta \dot{x}+\xi y=0
\end{equation*}%
or%
\begin{equation*}
\ddot{x}+\frac{\xi ^{2}}{4\eta ^{2}}x=0\text{ \ \ \ \ };\text{ \ \ \ \ }%
\ddot{y}+\frac{\xi ^{2}}{4\eta ^{2}}y=0.
\end{equation*}%
The solutions are

\begin{eqnarray*}
x &=&a\cos \left( \omega t\right) +b\sin \left( \omega t\right) \\
y &=&-b\cos \left( \omega t\right) +a\sin \left( \omega t\right) \\
p_{x} &=&\eta \left( b\cos \left( \omega t\right) -a\sin \left( \omega
t\right) \right) \\
p_{y} &=&\eta \left( a\cos \left( \omega t\right) +b\sin \left( \omega
t\right) \right)
\end{eqnarray*}%
with\ $\omega =\frac{\xi }{2\eta }.$ The Hamiltonian of this system
expressed in terms of the constants of integration is 
\begin{equation*}
H=\frac{\xi }{2}\left( a^{2}+b^{2}\right)
\end{equation*}%
and the Hamilton equations $\dot{x}=\{x,H\}$ and\ $\dot{y}=\{y,H\}$ give the
equalities 
\begin{eqnarray*}
-a\omega \sin \left( \omega t\right) +b\omega \cos \left( \omega t\right)
&=&\left\{ a,b\right\} b\xi \cos \left( \omega t\right) +\left\{ b,a\right\}
a\xi \sin \left( \omega t\right) \\
b\omega \sin \left( \omega t\right) +a\omega \cos \left( \omega t\right)
&=&-\left\{ b,a\right\} a\xi \cos \left( \omega t\right) +\left\{
a,b\right\} b\xi \sin \left( \omega t\right)
\end{eqnarray*}%
from which we easily read off the bracket between the constants of
integration $\left\{ a,b\right\} =\frac{\omega }{\xi }.$ From it we deduce
the brackets between the dynamical variables 
\begin{eqnarray*}
\left\{ x,y\right\} &=&-\frac{1}{qB_{0}} \\
\left\{ x,p_{x}\right\} &=&\left\{ y,p_{y}\right\} =\frac{1}{2} \\
\left\{ p_{x},p_{y}\right\} &=&\frac{1}{4}qB_{0}
\end{eqnarray*}

\subsubsection{Dirac approach}

We deduce from the same Lagrangian the following momenta $p_{x}=-\frac{qB_{0}%
}{2}y$, $p_{y}=\frac{qB_{0}}{2}x$ and the canonical Hamiltonian $H_{c}=\frac{%
\xi }{2}\left( x^{2}+y^{2}\right) .$ We therefore have two primary
constraints 
\begin{eqnarray*}
\phi _{1} &=&p_{x}+\frac{qB_{0}}{2}y\approx 0 \\
\phi _{2} &=&p_{y}-\frac{qB_{0}}{2}x\approx 0
\end{eqnarray*}%
The total Hamiltonian with the constraints is $H_{T}=H_{c}+\lambda _{1}\phi
_{1}+\lambda _{2}\phi _{2}.$ The consistency conditions give $\lambda
_{1}\approx -\frac{k}{B_{0}}y$ and $\lambda _{2}\approx \frac{k}{B_{0}}x$.\
We also see that these constraints are second class as $\left\{ \phi
_{1},\phi _{2}\right\} _{PB}=-\left\{ \phi _{2},\phi _{1}\right\}
_{PB}\approx qB_{0}\neq 0.$

The matrix of constraints is therefore

\begin{equation*}
M=\left( 
\begin{array}{cc}
0 & \left\{ \phi _{1},\phi _{2}\right\} _{PB} \\ 
\left\{ \phi _{2},\phi _{1}\right\} _{PB} & 0%
\end{array}%
\right) =qB_{0}\left( 
\begin{array}{cc}
0 & 1 \\ 
-1 & 0%
\end{array}%
\right)
\end{equation*}%
From this matrix we get the Dirac brackets of the system:%
\begin{eqnarray*}
\left\{ x,y\right\} _{D} &=&-\frac{1}{qB_{0}} \\
\left\{ x,p_{x}\right\} _{D} &=&\left\{ y,p_{y}\right\} _{D}=\frac{1}{2} \\
\left\{ p_{x},p_{y}\right\} _{D} &=&\frac{1}{4}qB_{0}
\end{eqnarray*}%
which is, as expected, are identical to some obtained with the CI method.

\subsubsection{Faddeev-Jackiw approach}

In this case the F-J Lagrangian is the same

\begin{equation*}
L_{FJ}=\eta \left( x\dot{y}-y\dot{x}\right) -\frac{1}{2}\xi \left(
x^{2}+y^{2}\right)
\end{equation*}%
with $p_{x}=-\eta y$ and $p_{y}=\eta x$, the Euler-Lagrange equations are
then%
\begin{eqnarray*}
\eta \dot{y} &=&-\eta \dot{y}+\xi x \\
-\eta \dot{x} &=&\eta \dot{x}+\xi y
\end{eqnarray*}%
or in matrix form%
\begin{equation*}
2\eta \underset{f}{\underbrace{\left( 
\begin{array}{cc}
0 & 1 \\ 
-1 & 0%
\end{array}%
\right) }}\left( 
\begin{array}{c}
\dot{x} \\ 
\dot{y}%
\end{array}%
\right) =\left( 
\begin{array}{c}
\xi x \\ 
\xi y%
\end{array}%
\right)
\end{equation*}%
The inverse matrix is%
\begin{equation*}
f^{-1}=\frac{1}{2\eta }\left( 
\begin{array}{cc}
0 & -1 \\ 
1 & 0%
\end{array}%
\right) =\left( 
\begin{array}{cc}
\left\{ x,x\right\} & \left\{ x,y\right\} \\ 
\left\{ y,x\right\} & \left\{ y,y\right\}%
\end{array}%
\right)
\end{equation*}%
and we again find the same brackets as with the two other approaches.

\subsection{Gauge invariant Lagrangian}

Consider now a system described by the Lagrangian 
\begin{equation*}
L=\frac{1}{2}(y\dot{x}+x\dot{y})^{2}.
\end{equation*}

\subsubsection{CI method}

The general solution of the equations of motion is obtained by introducing $%
Q=xy,$ and consequently $\dot{Q}=y\dot{x}+x\dot{y}.$ Then we have $\ddot{Q}%
=0 $ leading to the following results

\begin{eqnarray*}
x(t) &=&\left( at+b\right) \varepsilon (t)\ \ \ \ ;\ \ \ \ \ y(t)=\frac{1}{%
\varepsilon (t)} \\
p_{x}(t) &=&\frac{a}{\varepsilon (t)}\ \ \ \ ;\ \ \ \ p_{y}(t)=\left(
a^{2}t+ab\right) \varepsilon (t),
\end{eqnarray*}%
where $\varepsilon (t)$ is an arbitrary function of time. To determine the
infinitesimal gauge transformation that leaves the Lagrangian invariant we
consider an infinitesimal variation $\delta \varepsilon $. This leads to $%
\delta x=\left( at+b\right) \delta \varepsilon $ and $\delta y=-\frac{\delta
\varepsilon }{\varepsilon (t)^{2}}$ which can be written $\delta x=xy\delta
\varepsilon $ and $\delta y=-y^{2}\delta \varepsilon .$

Fixing the gauge with the condition $y-1=0\Rightarrow \varepsilon (t)=1$,
the solution becomes

\begin{eqnarray*}
\ x(t) &=&at+b\ \ \ \ \ ;\ \ \ \ \ \ y(t)=1 \\
p_{x}(t) &=&a\ \ \ \ ;\ \ \ \ p_{y}(t)=a^{2}t+ab.
\end{eqnarray*}%
Replacing this solution in $H=\dot{x}p_{x}+\dot{y}p_{y}-L,$ we obtain 
\begin{equation*}
H=\frac{1}{2}a^{2}
\end{equation*}%
and using Hamilton equation $\frac{dx}{dt}=\{x,H\}$ in order to have
directly the bracket 
\begin{equation*}
\{a,b\}=-1
\end{equation*}%
From this bracket among the constants, we determine the fundamental brackets
:%
\begin{eqnarray*}
\{x,y\} &=&\{y,p_{x}\}=\{y,p_{y}\}=0 \\
\{x,p_{x}\} &=&1 \\
\{x,p_{y}\} &=&x \\
\{p_{x},p_{y}\} &=&-p_{x}.
\end{eqnarray*}

\subsubsection{Dirac approach}

From the Lagrangian we can deduce the momenta $p_{x}=y(y\dot{x}+x\dot{y})$
and $p_{y}=x(y\dot{x}+x\dot{y})$. It is clear that we have a primary
constraint $\phi =xp_{x}-yp_{y}\approx 0.$ The canonical Hamiltonian is $%
H_{c}=\frac{1}{2}\frac{p_{x}^{2}}{y^{2}},$ and the total Hamiltonian is
defined as $H_{T}=H_{c}+\lambda \phi .$ The constraint is here first class
because it is the only constraint. In order to define the Dirac brackets, it
is known that one must add an additional condition fixing the gauge. Here we
choose the same as previously $\zeta =y-1=0$. Therefore the matrix of
constraints is 
\begin{equation*}
M=\left( 
\begin{tabular}{ll}
$\left\{ \phi ,\phi \right\} _{PB}$ & $\left\{ \phi ,\zeta \right\} _{PB}$
\\ 
$\left\{ \zeta ,\phi \right\} _{PB}$ & $\left\{ \zeta ,\zeta \right\} _{PB}$%
\end{tabular}%
\right) =\left( 
\begin{tabular}{ll}
$0$ & $y$ \\ 
$-y$ & $0$%
\end{tabular}%
\right) =\left( 
\begin{tabular}{ll}
$0$ & $1$ \\ 
$-1$ & $0$%
\end{tabular}%
\right) .
\end{equation*}%
Now we can calculate the Dirac brackets among the dynamical variables 
\begin{eqnarray*}
\left\{ x,y\right\} _{D} &=&\left\{ x,y\right\} _{PB}-\left\{ x,\phi
\right\} _{PB}\left\{ \phi ,\zeta \right\} _{PB}^{-1}\left\{ \zeta
,y\right\} _{PB}-\left\{ x,\zeta \right\} _{PB}\left\{ \zeta ,\phi \right\}
_{PB}^{-1}\left\{ \phi ,y\right\} _{PB}=0 \\
\left\{ x,p_{x}\right\} _{D} &=&\left\{ x,p_{x}\right\} _{PB}-\left\{ x,\phi
\right\} _{PB}\left\{ \phi ,\zeta \right\} _{PB}^{-1}\left\{ \zeta
,p_{x}\right\} _{PB}-\left\{ x,\zeta \right\} _{PB}\left\{ \zeta ,\phi
\right\} _{PB}^{-1}\left\{ \phi ,p_{x}\right\} _{PB}=1 \\
\left\{ x,p_{y}\right\} _{D} &=&\left\{ x,p_{y}\right\} _{PB}-\left\{ x,\phi
\right\} _{PB}\left\{ \phi ,\zeta \right\} _{PB}^{-1}\left\{ \zeta
,p_{y}\right\} _{PB}-\left\{ x,\zeta \right\} _{PB}\left\{ \zeta ,\phi
\right\} _{PB}^{-1}\left\{ \phi ,p_{y}\right\} _{PB}=x \\
\left\{ y,p_{x}\right\} _{D} &=&\left\{ y,p_{x}\right\} _{PB}-\left\{ y,\phi
\right\} _{PB}\left\{ \phi ,\zeta \right\} _{PB}^{-1}\left\{ \zeta
,p_{x}\right\} _{PB}-\left\{ y,\zeta \right\} _{PB}\left\{ \zeta ,\phi
\right\} _{PB}^{-1}\left\{ \phi ,p_{x}\right\} _{PB}=0 \\
\left\{ y,p_{y}\right\} _{D} &=&\left\{ y,p_{y}\right\} _{PB}-\left\{ y,\phi
\right\} _{PB}\left\{ \phi ,\zeta \right\} _{PB}^{-1}\left\{ \zeta
,p_{y}\right\} _{PB}-\left\{ y,\zeta \right\} _{PB}\left\{ \zeta ,\phi
\right\} _{PB}^{-1}\left\{ \phi ,p_{y}\right\} _{PB}=0 \\
\left\{ p_{x},p_{y}\right\} _{D} &=&\left\{ p_{x},p_{y}\right\}
_{PB}-\left\{ p_{x},\phi \right\} _{PB}\left\{ \phi ,\zeta \right\}
_{PB}^{-1}\left\{ \zeta ,p_{y}\right\} -\left\{ p_{x},\zeta \right\}
_{PB}\left\{ \zeta ,\phi \right\} _{PB}^{-1}\left\{ \phi ,p_{y}\right\}
_{PB}=-p_{x}.
\end{eqnarray*}%
These brackets are identical to those obtained with the CI method.

\subsubsection{Faddeev-Jackiw approach}

Using the primary constraint $xp_{x}-yp_{y}=0$, we define the FJ Lagrangian
as $L_{FJ}=\dot{x}p_{x}+\dot{y}p_{y}-H_{c}$ which reads

\begin{equation*}
L_{FJ}=p_{x}\dot{x}+\frac{xp_{x}}{y}\dot{y}-\frac{1}{2}\frac{p_{x}^{2}}{y^{2}%
}.
\end{equation*}%
The independent variables of the system are $x,$ $y,$ $z\equiv p_{x},$ then 
\begin{equation*}
L_{FJ}=z\dot{x}+\frac{xz}{y}\dot{y}-\frac{1}{2}\frac{z^{2}}{y^{2}}
\end{equation*}%
The Euler-Lagrange equations are

\begin{eqnarray*}
-\dot{z} &=&-\frac{z}{y}\dot{y} \\
-\frac{\dot{x}z}{y}+\frac{xz}{y^{2}}\dot{y}-\frac{x\dot{z}}{y} &=&\frac{xz}{%
y^{2}}\dot{y}-\frac{z^{2}}{y^{3}} \\
0 &=&-\dot{x}-\frac{x}{y}\dot{y}+\frac{z}{y^{2}}
\end{eqnarray*}%
which can also be written in the matrix form%
\begin{equation*}
\underset{f}{\underbrace{\left( 
\begin{array}{ccc}
0 & \frac{z}{y} & -1 \\ 
-\frac{z}{y} & 0 & -\frac{x}{y} \\ 
1 & \frac{x}{y} & 0%
\end{array}%
\right) }}\left( 
\begin{array}{c}
\dot{x} \\ 
\dot{y} \\ 
\dot{z}%
\end{array}%
\right) =\left( 
\begin{array}{c}
0 \\ 
-\frac{z^{2}}{y^{3}} \\ 
\frac{z}{y^{2}}%
\end{array}%
\right) .
\end{equation*}%
The matrix $f$ is antisymmetric and singular as $\det (f)=0$. The single
zero mode is $(%
\begin{array}{ccc}
-\frac{x}{z} & \frac{y}{z} & 1%
\end{array}%
)$ but it does not help as the zero mode equation does not lead to a
secondary constraint equation. The matrix $f$ is therefore still singular
and we are in the presence of a gauge symmetry. We choose the same gauge
condition $y=1$ and add the term $\dot{\lambda}(y-1)$ to the $L_{FJ}$
Lagrangian, i.e.;

\begin{equation*}
L_{FJ}=z\dot{x}+\frac{xz}{y}\dot{y}-\frac{1}{2}\frac{z^{2}}{y^{2}}+\dot{%
\lambda}(y-1)\text{ .}
\end{equation*}%
The Euler-Lagrange equations are now%
\begin{eqnarray*}
-\dot{z} &=&-\frac{z}{y}\dot{y} \\
-\frac{\dot{x}z}{y}+\frac{xz}{y^{2}}\dot{y}-\frac{x\dot{z}}{y} &=&\frac{xz}{%
y^{2}}\dot{y}-\frac{z^{2}}{y^{3}}-\dot{\lambda} \\
0 &=&-\dot{x}-\frac{x}{y}\dot{y}+\frac{z}{y^{2}} \\
-\dot{y} &=&0
\end{eqnarray*}%
or in the matrix form%
\begin{equation*}
\underset{f}{\underbrace{\left( 
\begin{array}{cccc}
0 & \frac{z}{y} & -1 & 0 \\ 
-\frac{z}{y} & 0 & -\frac{x}{y} & 1 \\ 
1 & \frac{x}{y} & 0 & 0 \\ 
0 & -1 & 0 & 0%
\end{array}%
\right) }}\left( 
\begin{array}{c}
\dot{x} \\ 
\dot{y} \\ 
\dot{z} \\ 
\dot{\lambda}%
\end{array}%
\right) =\left( 
\begin{array}{c}
0 \\ 
-\frac{z^{2}}{y^{3}} \\ 
\frac{z}{y^{2}} \\ 
0%
\end{array}%
\right) .
\end{equation*}%
The matrix $f$ is now antisymmetric and invertible because $\det (f)=1\neq 0$%
. The inverse matrix $f^{-1}$ is given by%
\begin{equation*}
f^{-1}=\left( 
\begin{array}{cccc}
0 & 0 & 1 & \frac{x}{y} \\ 
0 & 0 & 0 & -1 \\ 
-1 & 0 & 0 & -\frac{z}{y} \\ 
-\frac{x}{y} & 1 & \frac{z}{y} & 0%
\end{array}%
\right) =\left( 
\begin{array}{cccc}
\{x,x\} & \{x,y\} & \{x,z\} & \{x,\lambda \} \\ 
\{y,x\} & \{y,y\} & \{y,z\} & \{y,\lambda \} \\ 
\{z,x\} & \{z,y\} & \{z,z\} & \{z,\lambda \} \\ 
\{\lambda ,x\} & \{\lambda ,y\} & \{\lambda ,z\} & \{\lambda ,\lambda \}%
\end{array}%
\right) 
\end{equation*}%
and we find again%
\begin{eqnarray*}
\{x,y\} &=&\{y,z\}=0 \\
\{x,z\} &=&1
\end{eqnarray*}%
To obtain the brackets with $p_{y}$ one uses the primary constraint $p_{y}=%
\frac{xp_{x}}{y}$ and put $y=1$ at the end of the computation of the
brackets.

We can see on this particular example that the Dirac and Faddeev-Jackiw
approaches are closer to each other than the CI\ method.

\section{Hojman-Urrutia Lagrangian}

There are at least two physical motivations for the study of the theory
associated to this Lagrangian that we would like to point out briefly. The
first goes back to the publication of Hojman and Urrutia \cite{Hojman 1981}
where they built a Lagrangian from two systems of second order differential
equations for which in principle no second-order Lagrangian exists
(following the well known Douglas classification \cite{Douglas}).\ The
second motivation goes back to the work of Feynman, reported by Dyson \cite%
{Dyson} who derived the first group of Maxwell equations from the classical
equations of motion and imposed brackets among coordinates and velocities.
This work has been extended to the two groups of Maxwell equations \cite%
{Nous}. Although with this "Feynman brackets" approach, there is no
Lagrangian or Hamiltonian structure, Hojman and Shepley \cite{Hojman} showed
by using a Helmholtz inverse method, that under certain conditions a
Lagrangian can be associated to these Feynman brackets. In particular, in
order to test their method they studied the following equations of motion 
\begin{equation*}
\ddot{x}=-\dot{y}\text{ \ \ \ and \ \ \ }\ddot{y}=-y
\end{equation*}%
which are not derivable from a Lagrangian. By rewriting the system in first
order form as $\dot{x}=z,$ $\dot{y}=\omega ,$ $\dot{z}=-\omega $ and $\dot{%
\omega}=-y$ , they found the corresponding Lagrangian (by using integration
constants in another context)%
\begin{equation*}
L=(y+z)\overset{.}{x}+\omega \overset{.}{z}+\frac{1}{2}\left( \omega
^{2}-2yz-z^{2}\right) .
\end{equation*}%
This Lagrangian is singular and therefore this is a constrained system. This
theory has been studied via the Dirac approach in \cite{Kulshreshtha} and by
the Faddeev-Jackiw approach by \cite{Barcelos}. We do not reproduce their
results here but we will recover them from the CI method which turns out to
be much simpler.

\subsection{First case: $\protect\omega $ not constant}

The Euler-Lagrange equations are

\begin{equation*}
\begin{array}{c}
\dot{y}+\dot{z}=0 \\ 
0=\overset{.}{x}-z \\ 
\overset{.}{\omega }=\overset{.}{x}-y-z \\ 
0=\dot{z}+\omega%
\end{array}%
\text{ \ \ \ \ \ }\Leftrightarrow \text{ \ \ \ \ \ }%
\begin{array}{c}
\overset{.}{x}=z \\ 
\overset{.}{y}=\omega \\ 
\overset{.}{z}=-\omega \\ 
\overset{.}{\omega }=-y%
\end{array}%
\end{equation*}%
therefore $\overset{..}{y}=-y$ \ and \ $\overset{..}{x}=-\overset{.}{y\text{
(starting equations of Hojman and Shepley \cite{Hojman})}}.$

The general solution is

\begin{eqnarray*}
x &=&a\cos (t)+b\sin (t)+ct+d \\
y &=&-b\cos (t)+a\sin (t) \\
z &=&-a\sin (t)+b\cos (t)+c \\
\omega &=&a\cos (t)+b\sin (t)
\end{eqnarray*}%
and the Hamiltonian is%
\begin{equation*}
H=-\frac{1}{2}\left( a^{2}+b^{2}-c^{2}\right)
\end{equation*}%
From the Hamilton equation for $x$ we obtain the equality

\begin{eqnarray*}
-a\sin (t)+b\cos (t)+c &=&\left( -\left\{ a,b\right\} b+\left\{ a,c\right\}
c\right) \cos (t)+\left( \left\{ a,b\right\} a+\left\{ b,c\right\} c\right)
\sin (t) \\
&&+\left( \left\{ b,c\right\} b+\left\{ a,c\right\} a\right) t+\left(
\left\{ b,d\right\} b+\left\{ a,d\right\} a-\left\{ c,d\right\} \right) c
\end{eqnarray*}%
In the same manner for $y$

\begin{equation*}
b\sin (t)+a\cos (t)=\left( -\left\{ a,b\right\} a+\left\{ b,c\right\}
c\right) \cos (t)-\left( \left\{ a,b\right\} b+\left\{ a,c\right\} c\right)
\sin (t)
\end{equation*}%
which gives directly by identification%
\begin{eqnarray*}
\left\{ a,b\right\} &=&\left\{ c,d\right\} =-1 \\
\left\{ a,c\right\} &=&\left\{ a,d\right\} =\left\{ b,c\right\} =\left\{
b,d\right\} =0
\end{eqnarray*}%
and finally 
\begin{eqnarray*}
\left\{ x,y\right\} &=&\left\{ z,\omega \right\} =1 \\
\left\{ y,\omega \right\} &=&-1 \\
\left\{ x,z\right\} &=&\left\{ x,\omega \right\} =\left\{ y,z\right\} =0
\end{eqnarray*}%
Again, we obtain the same results as with the other methods. The reader will
notice the efficiency of our approach.

\subsection{Second case: $\protect\omega $ constant}

This case with $\omega =k$ constant studied also in \cite{Barcelos} is
interesting as it leads to a gauge theory \cite{Kulshreshtha} . In this
case, the Euler-Lagrange equations are%
\begin{eqnarray*}
\overset{.}{z} &=&0 \\
\overset{.}{x}-z &=&0 \\
\overset{.}{x}-z-y &=&0
\end{eqnarray*}%
whose general solution is

\begin{eqnarray*}
x &=&at+b \\
y &=&0 \\
z &=&a
\end{eqnarray*}%
The Hamiltonian is then

\begin{equation*}
H=\frac{1}{2}\left( k^{2}-2yz-z^{2}\right) =\frac{1}{2}a^{2}
\end{equation*}%
and from the Hamilton equation $\overset{.}{x}=\left\{ x,H\right\} $ we
obtain the equation $a=\left\{ b,\frac{1}{2}a^{2}\right\} $ which gives
directly the bracket among the constants 
\begin{equation*}
\left\{ a,b\right\} =-1
\end{equation*}%
From this we obtain the following brackets

\begin{eqnarray*}
\left\{ x,z\right\} &=&1 \\
\left\{ x,y\right\} &=&\left\{ y,z\right\} =0
\end{eqnarray*}%
as expected.

\section{Application to the Dirac field}

Here we present a different perspective on the quantization of free fields
based on the method we developed here. The idea is that, instead of applying
the quantization rules (for bosons and fermions) based on the correspondence
principle among the fields and their momenta \cite{Itzykson}, we simply
assume the validity of the Heisenberg equation of motion of the field
operators. Then we treat the creation and annihilation operators in the same
manner as the constants of motion of the classical equations of motion. As
an example we consider the Dirac field $\Psi $, with the Lagrangian density

\begin{equation*}
\mathcal{L}=i\overline{\Psi }\gamma ^{\mu }\partial _{\mu }\Psi -m\overline{%
\Psi }\Psi
\end{equation*}%
where $\gamma ^{\mu }$ are the Dirac matrices. The Euler-Lagrange equations
are $i\gamma ^{\mu }\partial _{\mu }\Psi $ $-m\Psi =0$ and $i\overline{\Psi }%
\gamma ^{\mu }\overleftarrow{\partial }_{\mu }$ $+m\overline{\Psi }=0$ whose
general solution is 
\begin{equation}
\Psi =\int \sum_{s=1}^{2}d\mathbf{k}\left( f_{k}(x)b_{s}(\mathbf{k})u_{s}(%
\mathbf{k})+f_{k}^{\ast }(x)d_{s}^{\dagger }(\mathbf{k})v_{s}(\mathbf{k}%
)\right)
\end{equation}%
where $f_{k}(x)=\sqrt{\frac{m}{(2\pi )^{3}k_{0}}}e^{-ikx}$, $k_{0}=\sqrt{%
\mathbf{k}^{2}+m^{2}}$ and $u_{s}(\mathbf{k)}$ and $v_{s}(\mathbf{k)}$ are
the usual bispinors. The $b_{s}(\mathbf{k})$ and $d_{s}(\mathbf{k})$ are
operators playing the role of the constants of motion in the classical case
(before quantization). Then as previously we write the Hamiltonian in terms
of these operators

\begin{equation}
H=\int d\mathbf{k}k_{0}\sum_{s=1}^{2}\left( b_{s}^{\dag }(\mathbf{k})b_{s}(%
\mathbf{k})-d_{s}(\mathbf{k})d_{s}^{\dagger }(\mathbf{k})\right)
\end{equation}

From the Heisenberg equation 
\begin{equation*}
\dot{\Psi}=\frac{1}{i\hslash }\left[ \Psi ,H\right]
\end{equation*}%
We obtain the following equalities 
\begin{eqnarray}
k_{0}b_{s}(\vec{k}) &=&\int d\vec{k}^{\prime }k_{0}^{\prime }\sum_{s^{\prime
}=1}^{2}\left[ b_{s}(\vec{k}),N_{s^{\prime }}(\vec{k}^{\prime })\right] \\
k_{0}d_{s}^{\dagger }(\vec{k}) &=&-\int d\vec{k}^{\prime }k_{0}^{\prime
}\sum_{s^{\prime }=1}^{2}\left[ d_{s}^{\dagger }(\vec{k}),N_{s^{\prime }}(%
\vec{k}^{\prime })\right]
\end{eqnarray}%
with $N_{s}(\vec{k})=\left( b_{s}^{\dag }(\mathbf{k})b_{s}(\mathbf{k})-d_{s}(%
\mathbf{k})d_{s}^{\dagger }(\mathbf{k})\right) .$ From these equations we
directly read off the following brackets 
\begin{equation}
\left[ b_{s}(\vec{k}),N_{s^{\prime }}(\vec{k}^{\prime })\right]
=b_{s^{\prime }}(\vec{k}^{\prime })\delta _{ss^{\prime }}\delta (\vec{k}-%
\vec{k}^{\prime })  \label{azertyuio}
\end{equation}%
\begin{equation}
\left[ d_{s}^{\dagger }(\vec{k}),N_{s^{\prime }}(\vec{k}^{\prime })\right]
=-d_{s^{\prime }}^{\dagger }(\vec{k}^{\prime })\delta _{ss^{\prime }}\delta (%
\vec{k}-\vec{k}^{\prime }).  \label{azertyuiop}
\end{equation}%
For fermions, we express these commutators by means of anticommutators%
\begin{eqnarray}
\left[ b_{s}(\vec{k}),N_{s^{\prime }}(\vec{k}^{\prime })\right]
&=&-b_{s^{\prime }}^{\dag }(\vec{k}^{\prime })\left\{ b_{s}(\vec{k}%
),b_{s^{\prime }}(\vec{k}^{\prime })\right\} +\left\{ b_{s}(\vec{k}%
),b_{s^{\prime }}^{\dag }(\vec{k}^{\prime })\right\} b_{s^{\prime }}(\vec{k}%
^{\prime })  \label{azertyuiop^} \\
&&+d_{s^{\prime }}(\vec{k}^{\prime })\left\{ b_{s}(\vec{k}),d_{s^{\prime
}}^{\dagger }(\vec{k}^{\prime })\right\} -\left\{ b_{s}(\vec{k}%
),d_{s^{\prime }}(\vec{k}^{\prime })\right\} d_{s^{\prime }}^{\dagger }(\vec{%
k}^{\prime })
\end{eqnarray}%
and%
\begin{eqnarray}
\left[ d_{s}^{\dagger }(\vec{k}),N_{s^{\prime }}(\vec{k}^{\prime })\right]
&=&-b_{s^{\prime }}^{\dag }(\vec{k}^{\prime })\left\{ d_{s}^{\dagger }(\vec{k%
}),b_{s^{\prime }}(\vec{k}^{\prime })\right\} +\left\{ d_{s}^{\dagger }(\vec{%
k}),b_{s^{\prime }}^{\dag }(\vec{k}^{\prime })\right\} b_{s^{\prime }}(\vec{k%
}^{\prime })  \label{azertyuiop^q} \\
&&+d_{s^{\prime }}(\vec{k}^{\prime })\left\{ d_{s}^{\dagger }(\vec{k}%
),d_{s^{\prime }}^{\dagger }(\vec{k}^{\prime })\right\} -\left\{
d_{s}^{\dagger }(\vec{k}),d_{s^{\prime }}(\vec{k}^{\prime })\right\}
d_{s^{\prime }}^{\dagger }(\vec{k}^{\prime }),
\end{eqnarray}%
then by identification of Eqs. (\ref{azertyuio}) and (\ref{azertyuiop^}) we
obtain the anticommutation rule%
\begin{equation}
\left\{ b_{s}(\vec{k}),b_{s^{\prime }}^{\dag }(\vec{k}^{\prime })\right\}
=\delta _{ss^{\prime }}\delta (\vec{k}-\vec{k}^{\prime })
\end{equation}%
In the same manner, the identification between Eqs. (\ref{azertyuiop}) and (%
\ref{azertyuiop^q}) gives%
\begin{equation}
\left\{ d_{s^{\prime }}(\vec{k}^{\prime }),d_{s}^{\dagger }(\vec{k})\right\}
=\delta _{ss^{\prime }}\delta (\vec{k}-\vec{k}^{\prime })
\end{equation}%
These rules of quantization are exactly identical to the results of the
canonical quantization of the Dirac field. The same approach can be used for
all quantum fields (Maxwell and Klein-Gordon).

\section{Conclusion}

The quantization of constrained systems is an old topic going back to Dirac
and Bergmann. Their approach with the introduction of first and second
class, primary and secondary constraints is essential but is very often
difficult to apply. Faddeev and Jackiw proposed an alternative approach
based on symplectic geometry and Darboux theorem which is very efficient and
generally simpler than Dirac approach. In this paper, we developed another
method that was applied on several examples and compared with the two other
approaches. This method requires the knowledge of the general solutions of
the equations of motion. This is for instance similar to the semiclassical
quantization from the path integral approach. It is practical because it is
very direct, easy and without special formalism. The novelty of this
approach is based on the (simple) computation of the (Dirac) brackets among
the integration constants (CI method) of the equations of motion. These
constants are therefore considered as variables, a method reminiscent of the
Hamilton-Jacobi calculus. We found that the CI method is consistent with the
two others and we believe that it is also simpler in general. It is however
based on the knowledge of the analytical solution, which is its weak point.
When the classical analytical solution is not known, the method can still be
applied by starting from the free systems because the solution is
accessible, and then use the perturbation method to go further. Even in the
others approaches, Dirac-Bergmann and Faddeev-Jackiw methods, one must find
the free solution after quantization in order to use the perturbation
development. In the near future we plan to extend our approach to non
soluble and more complicated constrained systems like spinning particles in
an electromagnetic field \cite{spinning} or relativistic particles in
non-commutative space-time \cite{doubly}. Beyond its technical aspect this
method sheds light also on another point of view of the quantization
process. Indeed, assuming only the validity of Hamilton (or Heisenberg)
equations for the quantum fields, the quantization rules emerge naturally
instead of postulating the commutation rules among the fields.

\bigskip

\textit{Acknowledgements. }We thank M. M. M\"{u}ller and Y. Grandati for
stimulating discussions. Z. B.\ thanks A. H. Gharbi for stimulating
discussions and for support.

\end{document}